\documentclass[aps,article,nofootinbib]{revtex4}

\usepackage{graphicx}

\usepackage{amsfonts}
\usepackage{amsmath}

\newcommand{\be}{\begin{eqnarray}}
\newcommand{\ee}{\end{eqnarray}}

\newcommand{\ox}{\overline{x}}
\begin{document}
\author{E.~Autieri}
\affiliation{Dipartimento di Fisica  Universit\`a degli Studi di Trento, Via Sommarive 14, Povo (Trento), I-38050 Italy.} 

\author{P.~Faccioli}
\affiliation{Dipartimento di Fisica  Universit\`a degli Studi di Trento e I.N.F.N, Via Sommarive 14, Povo (Trento), I-38050 Italy.} 
\email{faccioli@science.unitn.it}
\author{M.~Sega}
\affiliation{Frankfurt Institute for Advanced Studies, Ruth-Moufang-Str.~1, D-60435 Frankfurt, Germany}
\author{F.~Pederiva}
\affiliation{Dipartimento di Fisica and C.N.R./I.N.F.M.-DEMOCRITOS National Simulation Center,  Universit\`a degli Studi di Trento, Via Sommarive 14, Povo (Trento), I-38050 Italy.}
\author{ H.~Orland}
\affiliation{Institut de Physique Th\'eorique,
Centre d'Etudes de Saclay, CEA, IPhT, F-91191, Gif-sur-Yvette, France.}
\begin{abstract}
This  paper is devoted to the development of a theoretical and computational framework to  efficiently sample the statistically significant thermally activated reaction pathways, in multi-dimensional systems obeying Langevin dynamics.
We show how to obtain the set of most probable reaction pathways and  compute the corrections due to quadratic thermal fluctuations around such trajectories. We discuss  how to obtain predictions for the evolution of arbitrary observables and how to generate conformations which are representative of the transition state ensemble. We present an illustrative implementation of our method by studying the diffusion of a point particle in a 2-dimensional funneled external potential.
\end{abstract}

\title{Dominant Reaction Pathways in High Dimensional Systems}
\maketitle
\section{Introduction}

The problem of characterizing  thermally activated transitions in systems with a large number of degrees of freedom plays a pivotal role in the study of many physical phenomena. 
From a theoretical perspective, the most natural strategy to describe the dynamics of a generic system is to integrate numerically the equations of motion, e.g. via Molecular Dynamics (MD) simulations. In principle, once a statistically significant sample of MD transitions has been obtained, it is possible to derive predictions for the time evolution of the observables, which can be matched against experimental measurements.

Unfortunately, if the system contains a large number of degrees of freedom, or if the interesting transitions are rare events, characterized by free energy barriers of more than few units of $k_B~T$, MD approaches become extremely costly and often impracticable. This is essentially because most of  the computational time is invested in describing the motion of the system while it is exploring the portions of configuration space associated to the stable and meta-stable states. On the other hand, one is mostly interested in the information encoded in the reaction pathways joining the different states. 

A prototype of problem in which the MD approach is unfeasible is the study of the protein folding reaction. In this case, the typical number of atoms to be simulated is of order  $10^4$, the microscopic  time-scale associated with the motion of the torsional angles of the polypeptide chain is of the order of the $ns$, i.e. many orders of magnitude smaller than the macroscopic  time-scale associated with the inverse folding rate, which is in the range $ms-s$.
State-of-the-art all-atom MD simulations cover time intervals in the range $10-10^2~ns$\, (see e.g. \cite{SimAdv1} and references therein) and can  be useful only in investigating the reactions leading to local secondary structures formation, i.e.  $\alpha$-helices and $\beta$-sheets.

Given the limitations of the available MD simulations, several alternative approaches have been recently developed, to deal with the problem of characterizing the entire reaction in a high-dimensional space, using available computers. A drawback of these methods is that they rely either on  {\it ad-hoc} assumptions about the dynamics obeyed by the system \cite{Elber} or on a specific choice of reaction  coordinates \cite{TPS}.

In this work, we develop a theoretical framework which we call Dominant Reaction Pathways (DRP), which allows to rigorously identify the statistically significant transition pathways, without relying on any choice of reaction coordinates. Most importantly, the DRP approach is based on the low temperature expansion of the Langevin dynamics and, therefore, it does not involve any uncontrolled {\it ad hoc} approximation. 

A major advantage of the DRP approach is that it avoids investing computational time in simulating the local thermal motion in the metastable configurations. This is possible because the key equation can be formulated in a form which does not depend explicitly on the time variable. As a result, the computational difficulties associated with the existence of different time scales in rare thermally activated reactions, are naturally and rigorously bypassed. 

Some of the ideas underlying the DRP method were first introduced in ~\cite{DFP1, DFP2}, in the context of the study of the protein folding reaction. In those works, we showed that it is possible to  efficiently determine the {\it most probable} conformational transition pathways, both in the case of a coarse-grained and of a fully atomistic representation  of a molecule.
We now generalize this method to a much larger class of noise-driven phenomena in high-dimensional systems, and we present several theoretical developments.
In particular, we discuss how it is possible to obtain predictions for the evolution of arbitrary observables, while systematically including  the effect of thermal fluctuations around the most probable reaction paths. With this important development, the method does no longer focus on just the most probable transition trajectories in configuration space, but rather it allows to sample the entire ensemble of statistically relevant reaction pathways.

In the present paper we give a self-contained presentation of the theoretical formalism and of the relevant algorithms and we shall illustrate how these can be implemented in practice, by studying the simple and intuitive problem of the diffusion of a point-particle in a 2-dimensional external potential.
In a forthcoming publication, we shall present and test the application of the DRP method to the study of noise-driven conformational transitions of a polypeptide chain.

This paper is organized as follows. In section \ref{stochdiff} we review the path-integral formulation of stochastic Langevin diffusion and in \ref{obs} we relate it to the problem of computing the time evolution of experimentally observable quantities. In section \ref{important} we provide a presentation of low-temperature expansion underlying the DRP approach and provide a rigorous definition of the ensemble of statistically significant reaction pathways.
Sections \ref{DFP} and \ref{SQ} focus on the problem of efficiently sampling the ensemble of statistically significant reaction paths and  predicting the time evolution of the observables. The subsequent section 
\ref{tse} is devoted to the problem of generating an ensemble of configurations which is representative of the transition state ensemble. In section \ref{example} we provide a simple example of the application of the DRP method to the problem of the diffusion in a 2-dimensional funneled potential landscape. The main results and conclusions are summarized in \ref{conclusions}.

\section{Path Integral Formulation of the Langevin Dynamics} 
\label{stochdiff}

The starting point of our discussion of thermally activated transitions is the Langevin equation, which describes the out-of-equilibrium dynamics of a system coupled to a thermalized heat bath,
\begin{equation}
m~\ddot{x}=- \gamma \dot{x} - \nabla U(x) + \xi(t),
\label{Lan1}
\end{equation}
where $x=(x_1,x_2,...,x_N)$ is in general a vector specifying a point in configuration space  (for example, the coordinates of all the atoms in a molecule), $U(x)$ is a the potential energy and $\xi(t)$ is a memory-less Gaussian noise with zero average. 
For sake of simplicity and without loss of generality, in this section we  present the formalism in the case of one particle in one spacial dimension. The generalization to the multi-dimensional case is straightforward.

The term on the left hand side of  Eq. (\ref{Lan1}) introduces inertial effects which are damped on a time-scale $t\gtrsim m/\gamma\equiv \tau_D$.  In the following, we shall focus on physical systems for which the damping time scale $\tau_D$ is much smaller than the microscopic time-scale associated with the local conformational motion. For example,  in the case of the dynamics of a typical amino-acid chain,  $\tau_D$ is of  order of the fraction of a $ps$, hence much smaller than the relevant microscopic time-scale associated to the dynamics of the torsional angles, which takes place at the $ns$ time-scale.

For $t\gg\tau_D$ the Langevin Eq. (\ref{Lan1}) reduces to
\begin{equation}
\frac{\partial x}{\partial t} = -\frac{1}{k_B T} D~ \nabla U(x) + \eta(t),
\label{Lan2}
\end{equation}
where  $D=\frac{k_B T}{ m \gamma}$ is the diffusion coefficient which we shall assume to be independent on the particles position\footnote{For the sake of simplicity, we shall further assume that all the particles in the system have the same diffusion coefficient. The generalization to models in which each constituent has a different diffusion coefficient is trivial.} and $\eta(t)= \frac{1}{\gamma}~\xi(t)$ is a Gaussian noise that satisfy the fluctuation-dissipation relation,
\be
\langle\eta(t)\eta(t')\rangle= 2 D \delta(t-t'),
\ee
The probability distribution sampled by the stochastic differential Eq. (\ref{Lan2}) satisfies then the Fokker-Planck equation
\be
\frac{  \partial}{  \partial\,t}~P(x,t)= D \nabla 
\left(\frac{1}{k_B T}~\nabla\,U(x)~P(x,t)~\right)
+ D\,\nabla^2~P(x,t).
\label{FPE}
\ee
A universal property of all the solutions of the Eq. (\ref{FPE}) is that, in the long time limit, they converge to the Boltzmann weight, regardless of their initial conditions,
\be
P(x,t) \stackrel{(t\to\infty)}{\rightarrow} e^{-\frac{1}{k_B T} U(x)}.
\ee
Hence, the Fokker-Planck equation accounts for the approach to thermal equilibrium with the heat-bath.
By performing the substitution 
\be
P(x,t)= e^{-\frac{1}{2 k_B T} U(x)}~\psi(x,t)
\label{sub}
\ee
the Fokker-Planck Eq. (\ref{FPE}) can be recast in the form of a Schr\"odinger Equation in imaginary time:
\be
 -\frac{\partial}{\partial t} \psi(x,t) = \hat{H}_{eff}~\psi(x,t),
\label{SE}
\ee
where the effective "Quantum Hamiltonian" operator reads
\be
\hat{H}_{eff}~=~- D \nabla^2 + V_{eff}(x),
\label{Heff}
\ee
and the effective potential is defined as 
\be
V_{eff}(x)= \frac{D}{4 (k_B T)^2}~ \left[ \left(\nabla  U(x)\right)^2 - 2 k_B T\, \nabla^2 U(x)\right]
\label{Veff}
\ee
Hence, the problem of studying the diffusion of a classical particle with diffusion constant $D$  in a heat-bath at temperature $T$  can be mapped into the problem of determining the quantum-mechanical propagation in imaginary time of a virtual system, defined by the effective quantum Hamiltonian (\ref{Heff}), interacting with  the  effective potential $V_{eff}(x)$. 

Based on the analogy with quantum mechanics, it is immediate to obtain a path-integral representation of the solution of (\ref{FPE}), subject to the boundary conditions 
$x(-t/2)=x_i$ and $x(t/2)=x_f$:
\be
P\left(x_f,\frac{t}{2}\left|x_i,-\frac{t}{2}\right.\right) &=& e^{-\frac{1}{2 k_B T}(U(x_f)-U(x_i))} 
\langle x_i | e^{- \hat{H}_{eff} t}| x_f\rangle  \nonumber \\
&=& e^{-\frac{1}{2 k_B T}(U(x_f)-U(x_i))}~\int_{x_i}^{x_f} 
\mathcal{D}x(\tau)~e^{- \int_{-t/2}^{t/2} d\,\tau~ \left(\frac{\dot{x}^2(\tau)}{4 D}+ V_{eff}[x(\tau)]\right)}.
\label{path2}
\ee
This equation expresses the fact that the conditional Fokker-Planck probability is formally equivalent to a quantum-mechanical propagator in imaginary time.

In order to illustrate the physical content of the effective potential $V_{eff}(x)$ defined in Eq.(\ref{Veff}), let us consider the probability for the system to remain in the same configuration $x$, after an infinitesimal time-interval $dt$. One finds
\be
P(x, dt|  x, 0) \sim e^{-V_{eff}(x) dt}.
\ee 
Hence, the function $V_{eff}(x)$ measures the tendency of Langevin trajectories to evolve away from a given configuration $x$. In particular, the configurations where the effective potential is smallest  are those where a system will be trapped for the longest time. Conversely, the residence time will be shortest at configurations with the largest $V_{eff}$. Notice that such an effective interaction diverges in the low-temperature limit.
\begin{figure}[t!]
		\includegraphics[width=5 cm]{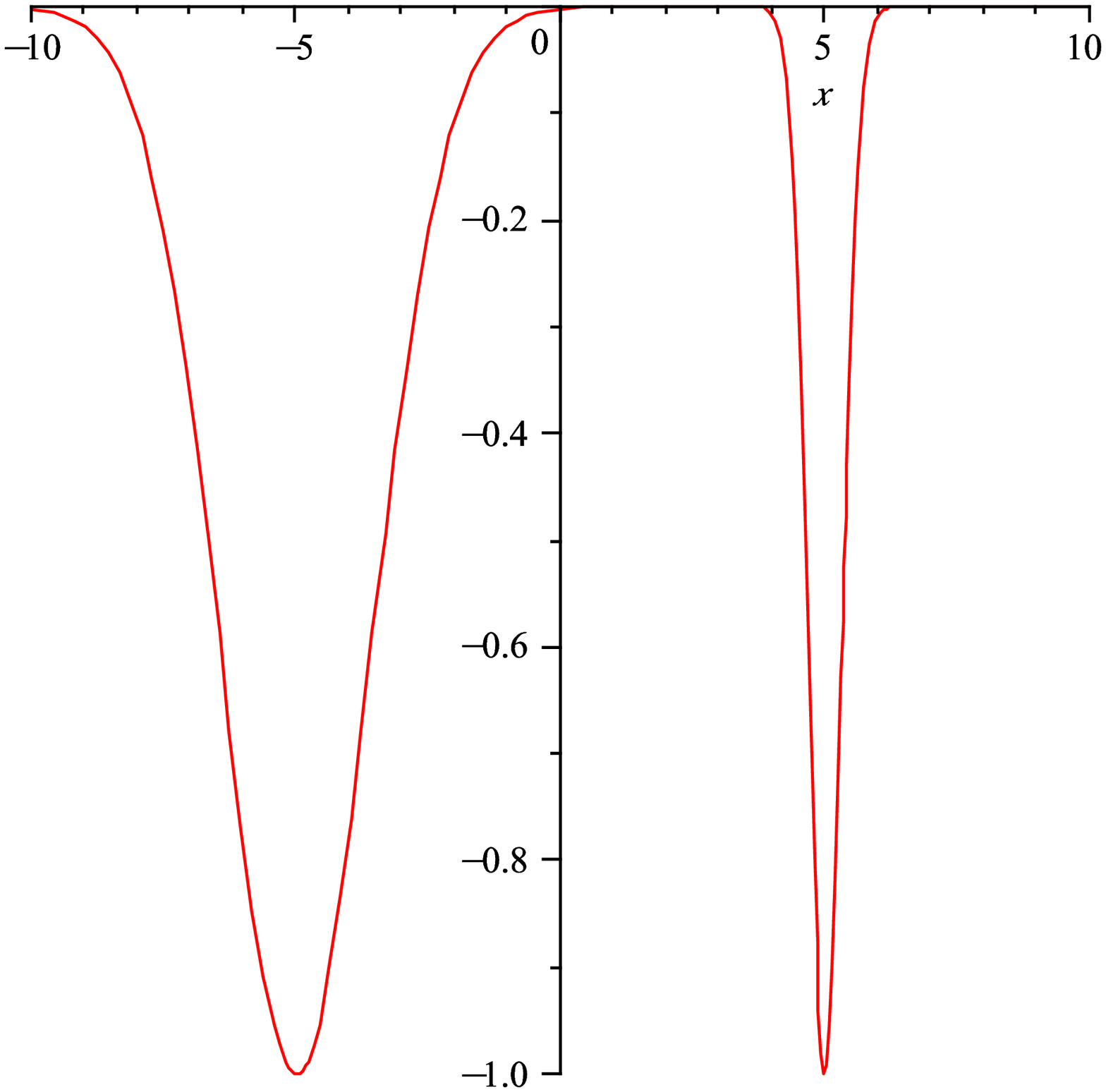}\qquad
				\includegraphics[width=5 cm]{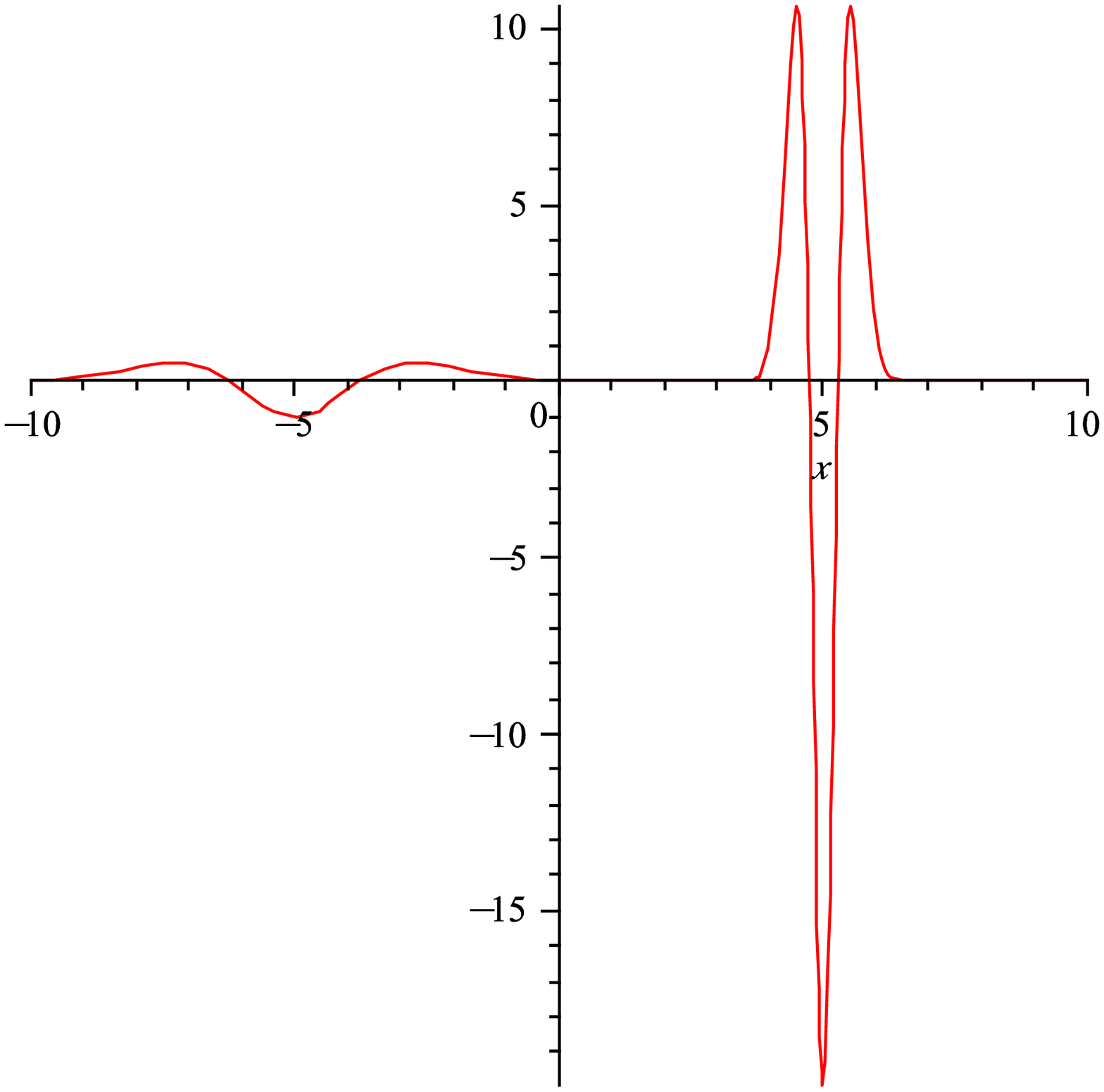}		
\caption{Left panel: an potential energy with two degenerate minima. Right panel: the corresponding effective potential $V_{eff}(x)$.}
\label{figexample}
\end{figure}
Such a property of the effective potential allows to interpret  physically  the two terms in Eq. (\ref{Veff}). The first term is always positive and provides the tendency of the system to move away from regions of configuration space where the forces are very large. In the low-temperature limit this is the driving effect. On the other hand, at higher temperatures, the interaction is dominated by the second term in (\ref{Veff}), which reduces the value of $V_{eff}$ in the vicinity of narrow minima of the physical potential $U$, i.e. when $\nabla^2 U$ is large and positive. This  implies that the system will be trapped for long times in regions of configuration space characterized by a small conformational entropy.  

These observations about the physical interpretation of $V_{eff}(x)$ are  illustrated by the example presented in Fig. \ref{figexample}. The left panel displays a potential function $U(x)$ characterized by two degenerate minima, the one on the left hand side being broader than that on the right hand side. The right panel displays the corresponding effective potential (obtained by setting both  $k_B T$ and  $D$ to 1). The configurations in the bottom of the broad minimum have a larger tendency to diffuse away, in an infinitesimal time, because there is a large number of nearby configurations which are thermally accessible. On the other hand, the configurations which are thermally accessible from the narrow minimum on the right are very few, because the potential raises very rapidly. As a result, the effective potential $V_{eff}(x)$ does not display degenerate minima, rather it has a pronounced global minimum in the proximity of the right minimum of $U(x)$. Note also that, the effective potential displays maxima where the forces are largest.


\section{Average time evolution of observables}
\label{obs}

In many reactions of interest, experiments cannot probe the kinetics of a single transition, but only measure the  {\it average} of a given quantity over a large number of independent transitions. 
For example, fluorescence experiments on the kinetics of  protein folding  measure the time evolution of the average relative distance between two specific amino-acid in the chain, during the 
reaction ---see e.g. \cite{Eaton} and references therein---.    If the system relaxes to thermal equilibrium, such quantities converge to their average value in the canonical ensemble. 
In this section we discuss how to relate the average time evolution of an observable to the path integral representation of the conditional Fokker-Planck probability (\ref{path2}). 
We do so by further exploiting the formal connection between the stochastic and the  
quantum dynamics.  

Let us begin by introducing an Hilbert space $\mathcal{H}$ spanned by the eigenvector of configuration operator
\be
\hat{X} |x \rangle = x |x \rangle.
\ee
Here $x$ denotes a point in the configuration space.
Let $O(x)$ be the  real-valued observable of interest, which we will assume to be a function of the system configuration only.  For example, $O(x)$ could be the position of a particle or the distance between two monomers in an amino-acid chain. We can  associate to $O(x)$  an operator $\hat O$,  acting on the Hilbert space $\mathcal{H}$:
\be
\hat{O} |x\rangle = O(x) | x \rangle,
\label{op}
\ee
We are now in a condition to  define the time-dependent average of the observable $\langle O(t)\rangle$:
\be
\langle \mathcal{O}(t') \rangle \equiv
\frac{\int dx e^{-\frac{U(x)-U(x_0)}{k_B T}} \langle x |\,e^{-\hat{H}_{eff} (t/2-t')}\, \hat{O} \,e^{-\hat{H}_{eff} t'} 
| x_0\rangle}{ 
\int dx e^{-\frac{U(x)-U(x_0)}{k_B T}} \langle x |\,e^{-\hat{H}_{eff} t} 
| x_0\rangle },
\label{Opdef}
\ee
where $t'$ is a arbitrary instant in the interval $-t/2 <t' <t/2$.
 In practice, one is mostly interested in  transitions which can take an arbitrary long time, i.e. those for which the total  time interval diverges,  $t \to\infty$.

A more physically transparent  representation of  (\ref{Opdef}) is obtained by inserting a  resolution of the identity, in terms of the eigenvalues of the configuration operator,~$| x'\rangle$:
\be
\langle \mathcal{O}(t') \rangle &=& 
\frac{ \int  dx\int dx' e^{-\frac{U(x)-U(x_0)-U(x')+U(x')}{2 k_B T}} \langle x |e^{-\hat{H}_{eff} (t/2-t')}\, | x' \rangle \langle x' | \hat{O} \,e^{-\hat{H}_{eff} (t'+t/2)} | x_0\rangle}
{  \int dx ~P(x, t/2| x_0, -t/2) }\nonumber\\
&=&
\frac{ \int dx' \int dx   P(x, t/2 | x' , t') O(x') P(x', t'| x_0, -t/2)  }
{  \int dx ~P(x, t/2| x_0, -t/2) }.
\label{Odef}
\ee

It is easy to check that this out-of-equilibrium average definition yields the correct result, in the small and large time limit.
In fact, if  the average is evaluated at initial time --- i.e. for  $t'\to (- t/2)$--- it returns the value of the observable in the initial configuration, i.e. 
\be
\lim_{t'\to-t/2} \langle \mathcal{O}(t') \rangle = O(x_0).
\ee

On the other hand, in appendix \ref{av} we prove that if  the average (\ref{Odef}) is evaluated in the  long time limit --- i.e. for  $t'\to (t/2)$ and $t\to \infty$---  it coincides with the equilibrium Boltzmann  average,
\be
\lim_{t'\to t/2, t \to \infty}   \langle O(t')\rangle =  \frac{\int dx e^{-\frac{U(x)}{k_B T}}\, O(x)}
{\int d x e^{-\frac{U(x)}{k_B T}}}.
\label{Obol}
\ee

The path integral representation of (\ref{Odef}) is immediately derived from the path integral representation of $P(x', t'|x_0, -t/2)$ and $P(x, t/2| x', t')$,  as in Eq. (\ref{path2}). We find 
\be
\langle O( t')\rangle  =
\frac
{
\int dx e^{-\frac{U(x)}{2 k_B T}} 
\int_{x_0}^{x(t)}  \mathcal{D} x(\tau)~O[x(t')]~e^{- \int_{-t/2}^{t/2} d\,\tau~ 
\left(\frac{\dot{x}^2(\tau)}{4 D}+ V_{eff}[x(\tau)]\right)}
}
{
\int dx e^{-\frac{U(x)}{2 k_B T}} \int_{x_0}^{x(t)}  \mathcal{D} x(\tau)~
e^{- \int_{-t/2}^{t/2} d\,\tau~ \left(\frac{\dot{x}^2(\tau)}{4 D}+ V_{eff}[x(\tau)]\right)}
}.
\label{GPI}
\ee

The discretized representation of (\ref{GPI}) is 
\be
\langle O( t')\rangle  &\equiv& \lim_{N\to \infty} \frac{ 
\int dx 
e^{-\frac{U(x)}{2k_B T}}
 \prod_{k=1}^{N-1} \left[ \int  dx_k \right]  O(x_{\hat{i}}) e^{- \frac t N \sum_k \frac{ (x_{k+1}-x_{k})^2 }{4 D (t/N)^2} +
 V_{eff}[x_{k}]}}
{ \int dx 
e^{-\frac{U(x)}{2 k_B T}}
\prod_{k=1}^{N-1} \left[ \int  dx_k \right]   e^{- \frac t N \sum_k \frac{ (x_{k+1}-x_{k})^2 }{4 D (t/N)^2} +
 V_{eff}[x_{k}]}
},
\label{dGPI}
\ee
where $i$ is the discretized time step closest to $t'$, i.e. $t' \simeq i \frac{t}{N}$.

Monte Carlo simulations offer in principle a scheme to compute the average (\ref{dGPI}). 
 If the interesting transitions are not rare events,  then the number of uniform time discretization steps  required to converge to the continuum limit is expected to be moderate.
On the other hand, if the interesting transitions are  rare, i.e. they occur at a time scale much longer than the microscopic time scale at which the local dynamics takes place, then the number of uniform discretization steps required to achieve the continuum limit would  diverge exponentially. In this regime, the direct evaluation of the average (\ref{GPI}) by Monte Carlo approaches rapidly becomes unfeasible.

In the following sections we shall discuss how, in some appropriate regime, the sampling of the most important reaction trajectories and the calculation of the evolution of the observables can be efficiently performed by applying the instanton calculus, a technique which was originally developed in the context of the functional integral description of quantum tunneling.

\section{Statistically Significant Reaction Pathways}
\label{important}

Owing to the stochastic nature of the underlying dynamics, thermally activated reactions always take place through different trajectories in configuration space. 
However, not all such trajectories can be considered as truly inequivalent reaction pathways. 
An ensemble made by an infinite number of paths will exhibit similar features. For example, during two transitions, the system may cross the same sequence of saddle points or, in the specific case of macro-molecules, may undergo the same pattern of contact formations.  Clearly, in these cases, it makes sense to view such reaction trajectories as different realizations of the same transition. On the contrary, trajectories which cross different saddles, or conformational transitions characterized by different sequences of contact formations represent truly distinct reaction pathways. In general, the distinct pathways will not all be equally probable and will therefore contribute differently to the reaction rate. 

These observations illustrate two main problems which emerge when attempting a characterization of the significant reaction pathways in high dimensional systems. The first consists in  providing a rigorous notion of reaction paths, i.e. one which does not rely on an a priori  choice of reaction coordinates.
The second problem regards the identification of the ensemble of most statistically relevant pathways, without having first to generate a huge sample of reaction trajectories by direct integration of the equation of motion. Once such a path has been identified, it is possible to make prediction for the average evolution of the observables and compare with experiments.

In this section,  we show how the path integral formulation of Langevin dynamics  provides a natural way  of defining, determining and characterizing the ensemble of most statistically significant reaction pathways, and to compute the time evolution of arbitrary observables, including the effect of thermal fluctuations, {\it without relying on any prior choice of reaction coordinate}.

We start from Eq. (\ref{path2}), which expresses the probability to make a transition from an initial configuration $x_i$ to a final configuration $x_f$, in terms of a sum over all possible transition pathways connecting these two points in configuration space.  
Since each path is assigned a well-defined probability, $e^{-S_{eff}[x]}$, the paths with the highest probability are those of minimum effective action $S_{eff}$, i.e. those satisfying the Euler-Lagrange equations of motion
\be
\frac{1}{2 D} \ddot x = \nabla V_{eff}(x),
\label{eqm}
\ee
with boundary conditions
\be
x\left(-\frac{t}{2}\right)= x_i, \qquad x\left(\frac{t}{2}\right)= x_f.
\label{bc}
\ee
These equations describe the classical (Newtonian) dynamics of a set of particles in the potential $-V_{eff}(x)$.
In the following, we shall refer to the solutions of (\ref{eqm}) obeying (\ref{bc}) as the dominant reaction trajectories.  They correspond to instantons in the language of quantum tunneling theory. We stress that, in order to correctly account for entropic and energetic effects it is crucial to include the Laplacian term in the effective potential. As it was clearly shown in \cite{adib},  approaches which do not account for the Laplacian contribution do not correctly identify the saddle-point pathways.

For a given choice of the boundary conditions, $x_i$ and $x_f$,  there are in general several dominant trajectories. Along with the exact solutions of the equation of motion associated to the effective action $S_{eff}$, there are also  {\it quasi}-exact solutions. For example, if the relevant transitions involve overcoming an energetic barrier, any combination of exact independent solutions of the equations of motion which cross the barrier back and forth several times represents itself a solution, up to terms which are linear in the temperature and exponentially small in the height of the barrier~\cite{Caroli2}. Such multiple-crossing trajectories correspond to multi instanton-antiinstanton solutions in the quantum mechanical language.

Obviously, since the exact and quasi-exact dominant trajectories form a functional set of zero measure, the probability to follow {\it exactly} any of them during a transition is zero. 
A significant contribution to the transition probability comes from the set of all the paths which lie in the functional vicinity of a given dominant trajectory and differ only by small thermal fluctuations. These paths form an ensemble of thermally equivalent realizations of the same transition.
By contrast, the statistical weight of the huge set of pathways which are not in the functional vicinity of a dominant trajectory is exponentially suppressed by the factor $e^{-S_{eff}}$. Hence, such paths do not provide a significant contribution to the path integral (\ref{path2})  and are very unlikely to occur.

Let us take a closer look on the parameter controlling the regime in which the contribution of the saddle-points in the path integral dominates over that of the stochastic fluctuations.
We begin by observing that the Fokker-Planck Eq. (\ref{FPE}) is invariant under a simultaneous rescaling of the diffusion coefficient and of the time variable, 
\be
D\to \alpha D\qquad t\to \frac 1 \alpha t.
\ee
Hence, without loss of generality, we can restrict our attention to the Fokker-Planck Eq. in the form 
\be
\frac{ \partial}{  \partial\,t}~P(x,t)=  \nabla 
\left(\nabla\,U(x)~P(x,t)~\right)
+ \theta\,\nabla^2~P(x,t).
\label{FPE2}
\ee
where $\theta$ is a parameter proportional to $k_B T$. 
The path integral representation of the Fokker-Planck equation  in the form (\ref{FPE2}) is
\be
P\left(x_f,\frac{t}{2}|x_i,-\frac{t}{2}\right)= e^{-\frac{1}{2 \theta}(U(x_f)-U(x_i))}\int_{x_i}^{x_f} \mathcal{D}x e^{-\frac{1}{2 \theta}\int d\tau  \frac{1}{2}  \dot{x}^2 +\frac 1 2  \left(
( \nabla U(x))^2 - 2 \theta \,\nabla^2 U(x)\right)},
\ee
Notice that in this expression the coefficient $2 \theta$ plays the role of the combination $\hbar/m$ in the  quantum mechanical path integral. Hence,  in general, the contribution of the paths in the vicinity of a saddle-point becomes important at low temperatures, when  the parameter  $\theta$ is small. This regime corresponds to the semi-classical limit, in the quantum language\,\cite{Caroli1}.

In the low temperature limit, it is possible to express the transition probability (\ref{path2}) as a sum over a countable set of contributions $i=1,2,...$, each of which accounts for the pathways which are close to a corresponding  dominant trajectory, denoted as $\ox_i(\tau)$. 
To obtain such a representation, let us take a completely general transition pathway $x(\tau)$ and re-write it as the sum of the i{\it-th} dominant  trajectory $\ox_i(\tau)$ and a fluctuation $y(\tau)$ around it:
\be
x(\tau)= \ox_i(\tau)+ y(\tau).
\ee
Note that, by construction, the fluctuation function $y(\tau)$ satisfies the boundary conditions $y(\tau/2)=y(-\tau/2)=0$.

We can now represent the contribution to the probability (\ref{path2}) of the paths near the i$-th$ dominant trajectory in the saddle-point approximation, i.e. by functionally expanding the action around  $\ox_i(\tau)$:
\be
S_{eff}[x] &=& S_{eff}[\ox_i]+ \frac{1}{2} \int_{-t/2}^{t/2} d\tau' \int_{-t/2}^{t/2} d\tau'' 
\frac{\delta^2 S_{eff}[\ox_i]}{\delta x(\tau') \delta x(\tau'')}~y(\tau)\,y(\tau) + \mathcal{O}(y^3)\nonumber\\
&\simeq& S_{eff}[\ox_i] + \frac{1}{2} \int_{-t/2}^{t/2} d\tau~y(\tau) ~\hat{F}[\ox]~ y(\tau),
\ee
where we have introduced the operator $\hat{F}$, defined as 
\be
\frac{\delta^2 S_{eff}[\ox_i]}{\delta x(\tau') \delta x(\tau'')} =
\left[-D \frac{d^2}{d \tau^2} + V_{eff}''[\ox_i]\right] \delta(\tau-\tau')\equiv ~\hat{F}[\ox_i] ~\delta(\tau-\tau'),
\ee
which governs the dynamics of the fluctuations around  $\ox_i$. Note that the expansion of the action does not contain linear terms in the fluctuation field $y(\tau)$, because the  path around which we expand is a solution of the equations of motion. 

We now expand the fluctuation function $y(\tau)$ in the basis of the (real) complete set of eigenfunctions of $\hat{F}$:
\be
y(\tau)= \sum_n c_n x_n(\tau), \qquad
\hat{F}[\ox_i] x_n(\tau) = \lambda_n x_n(\tau).
\ee
where the eigenfunctions $x_n$ are  chosen so as to satisfy the boundary conditions
$x_n(\pm t/2)=0$ and the normalization condition 
\be
\int_{-t/2}^{t/2} d\tau x_m(\tau) x_n(\tau) = \delta_{m n} 
\ee
To second order in the thermal fluctuations, the action reads
\be
S_{eff}[x] = S_{eff}[\ox_i]+ \frac{1}{2}\sum_n c_n \lambda_n^2 + ...
\ee
and the measure of the functional integral can be re-written as
\be
\mathcal{D} x = \mathcal{D} y =  \int_{-\infty}^{\infty}~\prod_n~\frac{d c_n}{ \sqrt{(2\pi)^d}}. 
\ee
With this choice, the contribution of the thermal fluctuations around the $i-th$ dominant trajectory to the
 transition probability can be finally written as
\be
P_i(x_f,t/2|x_i,t/2)&\simeq& e^{-\frac{1}{2 k_B T}(U(x_f)-U(x_i))}~e^{-S_{eff}[\ox_i]}\int_{-\infty}^{\infty}~\prod_n \frac{d c_n}{ \sqrt{(2\pi)^d}}
e^{- \lambda_n^2 c_n}  \nonumber \\
&=&  \frac{e^{-\frac{1}{2 k_B T}(U(x_f)-U(x_i))}}{\sqrt{\det \hat{F}[\ox_i]}}~e^{-S_{eff}[\ox_i]}.
\label{KtWKB}
\ee

The formal expression (\ref{KtWKB}) represents a reliable estimate of the contribution of the $i-th$ dominant trajectory to the transition probability (\ref{path2}) only if the total time interval $t$ is chosen to be of the same order magnitude as the microscopic time associated to the local motion of the relevant degrees of freedom $x$. 
 
On the other hand, in the study of thermally activated reactions, 
we are interested in transitions which take place on a time scale much larger than  the microscopic time scales.
Under such circumstances, the prefactor in the expression (\ref{KtWKB}) diverges and the approximate expression in Eq.~(\ref{KtWKB}) is no longer accurate, and has to be corrected~\cite{Caroli2}.

 To see how this problem emerges, it is useful to analyze the behavior of  (\ref{KtWKB}) in the limit in which the time interval for the transition diverges  $t\to\infty$. In such a limit, the  time-translation invariance of the solutions of Eq. (\ref{eqm})  is restored, hence 
 denoting by $t_0$ the time at which a transition occurs  one has 
 \be 
 \ox_i(t,t_0)=\ox_i(t-t_0).
 \ee 
Let us now consider the function
\be
x_0(\tau-t_0)= \frac{1}{2D S_{eff}[\ox_i]} \dot{\ox_i}(\tau-t_0)
\ee
It is immediate to verify that this solution is an eigenstate of the fluctuation operator with zero eigenvalue (zero-mode). To this end it is sufficient to take the derivative of the equations of motion:
\be
\ddot \ox_i - 2 D V_{eff}(\ox_i)  = 0
\ee
which leads to
\be
0 = (- 1/2D \partial^2_\tau + V_{eff}) x_0(\tau) = -\hat{F}[\ox_i] x_0 
\ee

The physical origin of this zero-mode resides in the fact that, in  the large $t$ limit, the functional minimum of the action does not reduce to a single path;  indeed, there is an infinite family of degenerate paths. Physically, such a large degeneracy reflects the fact that the crossing of the barrier can take place at any time between $-\infty$ and $\infty$ and that single-crossing events  can occur with equal probability at any time.

In the presence of a zero mode, the integral over $c_0$ in (\ref{KtWKB}) is divergent and  cannot be performed in the Gaussian approximation. This problem has been circumvented by replacing the integral over $c_0$ with an integral over the time at which the transition takes place (see e.g. \cite{zinn}). It is easy to show (see Appendix \ref{dc0}) that the integral over $d c_0$ can be related to the time at which the transition takes place,  
\be
d c_0 = \sqrt{2 D S_{eff}[\ox_i]} d t_0.
\ee 
Thus, the integration over $d c_0$ can be performed exactly, leading to our final result
\be
P_i(x_f,t/2|x_i,t/2)\simeq  
t ~A_i~e^{-\frac{1}{2 k_B T}(U(x_f)-U(x_i))}~e^{-S_{eff}[\ox_i]},
\label{KtWKB2}
\ee
where 
\be
A_i=\sqrt{\frac{2 D S_{eff}[\ox_i]}{\det' \hat{F}[\ox_i]}},
\label{Ai}
\ee
and the symbol $\det'\hat{F} $ denotes the product of all strictly positive eigenvalues of the fluctuation operator.

The complete representation of the transition probability $P(x_f,t/2|x_i,t/2)$ in the saddle-point approximation  is obtained by summing up the contributions associated to the functional vicinity of each minimum of the $S_{eff}$ action, i.e.  
\be
P(x_f,t/2|x_i,-t/2)\simeq\sum_i P_i(x_f,t/2|x_i,-t/2).
\label{sumPi}
\ee

We stress the fact that each of the $P_i$ in Eq. (\ref{sumPi}) includes the contribution of an ensemble of reaction paths, which are close to the same dominant reaction path. Hence, in the regime in which the saddle-point approximation is reliable, each dominant reaction path can be considered as the representative element of an infinite set  of reaction pathways, which are equivalent up to small thermal fluctuations. Different  dominant trajectories represent inequivalent reaction pathways and the set of all such inequivalent paths characterizes the ensemble of all possible distinct and statistically relevant pathways through which the transition can occur. To lowest-order in the saddle-point approximation, an estimate for the relative probability of each class of reaction pathway is  provided by the weight $\exp[-S^{i}_{eff}]$.

In the remaining part of this section, we discuss how to use the saddle-point approximation to calculate the  average time-evolution of an arbitrary observable, over a statistically significant ensemble of independent transitions.

According to the definition (\ref{Odef}), one needs to average over the contribution of the dominant reaction pathways connecting the same initial configuration $x_0$ to an ensemble of final configurations distributed according to $e^{-\frac{U(x)}{2 k_B T}}$. 
To lowest-order, it is immediate to obtain the  saddle-point expression for $\langle O(t')\rangle$, as it amounts to retaining only the contribution of  the dominant pathways:
\be
\langle O(t') \rangle  \simeq \frac {\int dx e^{-\frac{U(x)}{2 k_B T}}~\sum_i O[\ox_i(t')]~e^{-S_{eff}[\ox_i; x,x_0,t]}}{\int dx e^{-\frac{U(x)}{2 k_B T}}~ {\sum_i ~e^{-S_{eff}[\ox_i; x,x_0,t]}}},
\label{GLO}
\ee
where the sum runs over the  dominant paths $\ox_i(\tau)$, which  connect $x_0$ to some $x$ 
in a time interval $t$.

At the  next-to-leading order in the saddle-point approximation, each exponential weight in the sum (\ref{GLO}) is multiplied by the pre-factors $A_i$, representing the measure of the space of thermal fluctuations
around each saddle-point,
\be
\langle O(t') \rangle \simeq \frac {\int dx e^{-\frac{U(x)}{2 k_B T}}~ \sum_i  O[\ox_i(t')]~A_i~e^{-S_{eff}[\ox_i; x,x_0,t]}}{{\int dx e^{-\frac{U(x)}{2 k_B T}}~ \sum_i A_i~e^{-S_{eff}[\ox_i; x,x_0,t]}}}.
\ee 
In practice,  computing explicitly the $A_i$ coefficients is a very challenging task, except for very simple  systems (see e.g. \cite{Caroli2}). In addition,  if the initial configuration $x_0$ is not completely determined,  one needs to perform an average over its probability distribution. 

The next sections are devoted to the problem of determining numerically the ensemble of most statistically significant transition pathways and calculating the average time-evolution of reaction coordinates, using numerical techniques. 

\section{Numerical Determination of the Dominant Reaction  Pathways}
\label{DFP}

In order to see how the dominant reaction pathways can be numerically determined, let us begin by recalling that each of them is a solution of the equation of motion 
\be
\frac{1}{2D} \ddot \ox (\tau)= \nabla V_{eff}(\ox).
\label{DFPeom}
\ee
The solutions of  (\ref{DFPeom}) conserve an effective "energy" $E_{eff}$ defined as
\be
E_{eff}= \frac{1}{4D} \dot{\ox}^2 - V_{eff}(\ox).
\ee
Note that $E_{eff}$  has the dimension of a rate and therefore does not have the physical interpretation of a mechanical energy.

As a consequence of the conservation of $E_{eff}$, the symplectic  action $S_{eff}$ evaluated along a dominant trajectory  can be rewritten as 
\be
S_{eff}(x_f,x_i, t) = - E(t) t +  S_{HJ}(x_f,x_i; E_{eff}(t)),
\ee
where $S_{HJ}(x_f, x_i; E_{eff})$ is the Hamilton-Jacobi (HJ) action defined as
\be
S_{HJ}(x_f, x_i;E_{eff}(t)) = \frac{1}{\sqrt{D}}~\int_{x_i}^{x_f}d l \sqrt{E_{eff}(t) + V_{eff}[\ox(l)]},
\label{SHJ}
\ee
where $dl = \sqrt{d x^2}$ is a measure of the distance covered by each constituent during the transition. 
This action can be seen to be identical to the argument of the exponent in  the wave-function, in the semi-classical WKB approximation of quantum mechanics.

We stress the fact that, in general,  the numerical value of the effective energy $E_{eff}$ depends on the choice of  total time of the transition time through the relationship
\be
t= \int_{x_i}^{x_f} dl ~\frac{1}{\sqrt{4 D~(E_{eff}(t)+V_{eff}[\ox(l)])}}.
\label{time}
\ee

Eq. (\ref{SHJ}) expresses the fact that the dominant trajectory in configuration space leaving from $x_i$ and reaching $x_f$ is a minimum of the HJ functional. Note that, at finite temperature, the minimum of the physical potential $U(x)$ does not in general coincide with the minimum of the effective potential $V_{eff}(x)$. However, the displacement between the minimum of $U(x)$ and of  $V_{eff}(x)$ are of order $k_B T$.    

In  \cite{DFP2} we have shown that, in the context of the study of configurational transitions of macro-molecules, the possibility of adopting the HJ formulation of the dynamics  leads to an impressive computational simplification of the problem. In fact, the total distance between two different molecular configurations is at most 1-2 orders of magnitude larger than the most microscopic length scale, i.e. the typical monomer (or atom) size. As a consequence, $\sim 50-100$ discretized displacement steps are usually sufficient for achieving convergence in the discretization of the trajectories. This number should be compared with $10^{12}$ time-steps which would be required to characterize a protein folding trajectory adopting the time-dependent formulation.

Once a dominant trajectory $\ox(l)$ has been determined, one can recover the information on the time evolution of the system during the transition by means of the well-known relationship
\be
\tau(x)= \int_{x_N}^{x} dl ~\frac{1}{\sqrt{4 D~(E_{eff}(t)+V_{eff}[\ox(l)])}},
\label{time2}
\ee
where $\tau(x)$ is the instant between $-t/2$ and $t/2$ at which the path visits the configuration $x$. Hence, through Eq. (\ref{time2}) it is possible to identify the configurations which are most probabily visited at each instant of time, i.e. the function $\ox(\tau)$.

Let us now discuss the implications of the choice of the effective energy, $E_{eff}$. From Eq. (\ref{time2}) it follows immediately that large effective energies are associated to short transition times. In addition, from Eq. (\ref{SHJ}) we see that in the infinite effective energy limit, trajectories become insensitive on the structure of the interaction and the dominant path is the straight line in the multi-dimensional configuration space.

In thermally activated processes, we are interested in transitions which take place on a very long time scale, which decouples from the microscopic time scale associated to the local motion in the stable and meta-stable states. In a protein folding transition, the final state $\ox(t)$ should be taken as the native state of the protein $\ox_f$, that is the absolute minimum of $U(x)$, or possibly a minimum of $V_{eff}(x)$ close to it. In order for the protein to stay in this state as long as possible, the velocity of the system at that point should be $\dot \ox(t) = \dot \ox_f =0$, which implies
 \be
E_{eff}= \frac{1}{4D}\dot{\ox}^2(t)-V_{eff}(\ox(t))=-V_{eff}(\ox(t)).
\label{Efft}
\ee


In conclusion, we have found that in thermally activated rare transitions, the most probable reaction pathways  are those which minimize the functional
\be
S_{HJ}(x_f, x_i;-V_{eff}[x_f]) = \frac{1}{\sqrt{D}}~\int_{x_i}^{x_f} d l 
\sqrt{ V_{eff}[\ox(l)]-V_{eff}[x_f]},
\label{SHJ2}
\ee
 In appendix \ref{free} we present a simple analytic example of the evaluation of the estimate for the transition probability, to lowest order in the saddle-point approximation, using the HJ formalism.
 
The numerical determination of the dominant reaction pathways involves searching for the minima of the discretized version of HJ functional (\ref{SHJ}). For example, if we are interested in molecular transitions, the discretized HJ functional  reads 
\be
S_{HJ}=\frac{1}{\sqrt{D}}~\sum_n^{N-1}\sqrt{\left(V_{eff}(n)-V_{eff}(1)\right)} \Delta l_{n,n+1}+\lambda P,
\label{DSHJ}
\ee 
where  $N$ is the total number of discretization steps and
\be
V_{eff}(n)&=&\frac{D}{4 (k_BT)^2} \sum_{i=1}^{N} \left[  \left(\sum_{j=1}^{N} {\bf \nabla}_j U({\bf x}_1(n),...,{\bf x}_N(n))\right)^2 -2 k_B T \sum_{j=1}^{N_p} \nabla^2_j U({\bf x}_1(n),...,{\bf x}_N(n))\right]\\
P&=&\sum_n^{N-1} (\Delta l_{n,n+1}-\langle \Delta l\rangle)^2,\\
(\Delta l)^2_{n,n+1}&=&\sum_{i=1}^{N_p}({\bf x}_i(n+1)-{\bf x}_i(n))^2.
\ee
$N_p$ is the number of particles (atoms  or residues) in the molecule, $\Delta\,l_{n,n+1}$ is the Euclidean measure of the $n-th$ elementary path-step and
$\lambda~P$ is a penalty function which keeps all the length elements close to their average~\cite{Elber}. Although this term becomes irrelevant in the continuum limit, it plays an  important role if the transitions are rare events. In fact, it enforces that the sampling is made at {\it constant displacement steps} rather than at constant {\it time steps}.
The minimization of the discretized HJ functional (\ref{DSHJ}) can be performed for example by means of a simulated annealing algorithm.

We conclude this section by noting that the coordinate $l$ appearing in (\ref{SHJ}) can be interpreted as a reaction-coordinate. However, we stress that this quantity is  not chosen {\it a priori}, but rather emerges naturally and in a self-consistent way from the path-integral formulation of the Langevin dynamics.
The dominant path encodes some information about the structure of the free energy landscape along such a coordinate. In fact, one expects that the residence time in a configuration visited by a dominant path to be largest (smallest) in regions where the free-energy is smallest (largest). 
Indeed, numerical simulations of the conformational transitions of alanine dipeptide fully supported this expectation \cite{DFP2} and showed that this way it was possible to locate very accurately the maxima (minima) of the free energy landscape.

\section{Thermal Fluctuations around the Dominant Reaction Pathways}
\label{SQ}

In the previous section we have shown how to determine the set of dominant reaction pathways corresponding to a given choice of boundary conditions. 
In this section, we show how  it is possible to include the effect of quadratic thermal fluctuations around a dominant path,  by means of the Monte Carlo algorithm. For the sake of definiteness, we focus on molecular transitions. The method, however, holds in general for thermally activated reactions in arbitrary systems obeying Langevin dynamics.

Let us denote by
\be
\ox(n)=({\bf \ox}_1(n),...,{\bf \ox}_{N_p}(n)),
\label{traj}
\ee
with $n=1,..,N$,  the dominant trajectory corresponding to a sequence of configurations of the molecule, and determined by minimizing numerically the HJ action (\ref{DSHJ}). 

The time at which each configuration $\ox(n)$ is visited during the transition can be obtained  by computing the set of time intervals separating each of the path steps $\ox(n)$ from $\ox(n+1)$:
\be
\Delta \tau_{n,n+1}= \frac{\Delta l_{n, n+1}}{\sqrt{4D (V_{eff}(\ox(n))-V_{eff}(\ox(1)))}}
\ee
For example, the $10-th$ configuration $\ox(10)$ is visited at the time $\tau(10)= \sum_{n=1}^{10-1} \Delta \tau_{n,n+1}$. 

The sequence of time intervals obtained this way can be used to write a discretized version of the transition probability (\ref{path2}):
\be
P(x_f,x_i;\tau(N)) = \int~\prod_{n=1}^{N-1} \left[ d{\bf x}_1(n)... d{\bf x}_{N_p}(n)\right]~ e^{-\sum_{n=1}^{N-1} \Delta \tau_{n,n+1} ~\left[\frac{1}{4 D}~\sum_{i=1}^{N_p}~\left( \frac{{\bf x}_i(n+1)-{\bf x}_i(n)}{\Delta \tau_{n, n+1}}\right)^2 + V_{eff}({\bf x}_1(n), ..., {\bf x}_{N_p}(n))\right]}
\label{DPI}
\ee

Eq. (\ref{DPI}) represents a particularly convenient Trotter discretization of the original path integral (\ref{path2}). In fact, the time intervals are chosen large, when the most probable trajectory at that time is evolving slowly. On the other hand, the time intervals become very small when the configurations in the dominant path are moving fast. Such a construction is reliable only in the low temperature limit, where the information encoded in the saddle-point is relevant. In this regime, the number of Trotter discretization steps  $N$  is the same as the number of {\it path} discretization steps in (\ref{DSHJ})  and is much smaller than the number of equal time-discretization steps, which would be required if one had no prior knowledge of the relevant transition.   From the computational point of view, the sampling of the ensemble of fluctuations around each saddle-point in (\ref{DPI}) with a Monte Carlo algorithm is essentially as expensive as finding the dominant path itself. In fact, in both case one has to sample a space with $3 N_p\times N$  degrees of freedom.

 In order to single out the next-to-leading corrections in the saddle-point approximation of the propagator, one should in principle expand the action to quadratic order around the saddle-point first and then perform the MC sampling. However, in the regime of validity of the saddle-point approximation,   the contribution to the propagator of the cubic, quartic and higher terms in the expansion of the action are small and lead to corrections which come in at next-to-next-to-leading order. Hence, in such a regime, one does not need to perform the expansion of the action. On the other hand, if this condition is not satisfied, i.e. if thermal fluctuations are large, then the entire saddle-point approach breaks down.
The approximation scheme presented in this section can be applied also to derive estimate for the average value of observable, at next-to-leading order.

\section{Characterization of the Transition State Ensemble}
\label{tse}

The DRP formalism provides a prescription to identify a set of conformations which can be taken as the most representative of the Transition State Ensemble (TSE). 

To see how is it possible to extract such information from the dominant reaction pathways, we adopt the definition TSE in terms of commitment analysis~\cite{bunsen,pandeTS},  i.e. as the ensemble of conformations from which the probability to evolve into the native state equals the probability of collapsing into a denatured conformation.
Once a dominant path has been determined, it is immediate to identify the conformation $x_{ts}$ such that the probability in the saddle-point approximation
to diffuse back  to the initial configuration $x_i$ 
 equates the probability of evolving toward the final configuration $x_f$, i.e.
 \be
 P(x_N|x_{ts})=P(x_D|x_{ts}).
 \ee
 
To lowest order in the saddle-point approximation, this condition leads to  the simple equation:
\be
\frac{U(x_i)-U(x_f)} {2 k_B T } = S_{HJ}([x];x_{ts},x_f)-S_{HJ}([x];x_{ts},x_i).
\ee

The determination of the representative elements of the TSE can be generalized beyond the lowest-order, of the saddle-point approximation. To this end, we recall that the dominant reaction pathway is obtained in practice by minimizing numerically  the discretized version of the HJ action (\ref{DSHJ}). Hence, the configuration $x_{ts}$ which is most representative of the TSE corresponds to a given path step, $n_{ts}$. 
Once the most significant thermal fluctuations around the dominant pathway have been determined, using the prescription described in section \ref{SQ}, it is possible to determine a full ensemble of representative configurations of the transition state,  
\be
TSE \supset \{x_1(n_{ts}), ...,x_N(n_{ts})\},
\ee 
where the index $1..N$ labels all the fluctuation paths, and $x_i(n)$ with $i=1,..N$ are the paths determined with Monte Carlo sampling of the integral (\ref{DPI}).

\section{An Illustrative Example: Noise-Driven Escape from a 2-Dimensional Funnel}
\label{example}
\begin{figure}[t!]
		\includegraphics[width= 10 cm]{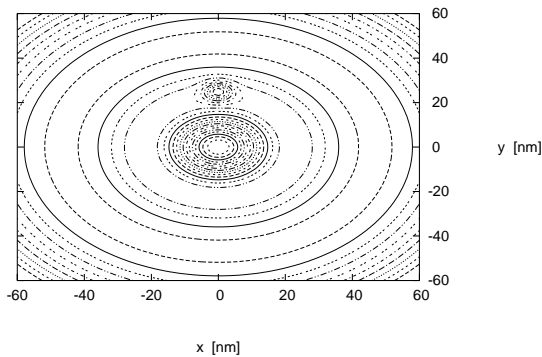}		
			\caption{Two dimensional energy landscapes considered in section \ref{example}.}
\label{landscape}
	\end{figure}

In view of applications of the DRP method to the study of  transitions which take place in complex high-dimensional spaces, it is instructive to first illustrate how this approach can be implemented, in a very simple and intuitive case. In such a perspective, in this section we consider a 2-dimensional model, which  displays some of the key features of the protein folding reaction.  We  consider a funneled energy function $U({\bf x})$, whose minimum is narrow and several $k_B T$ deep. Such a minimum is separated by a flat circular region by an energetic barrier of some $k_B T$. The flat region is corrugated by adding a hump or a hole, of few  $k_B T$.  The system is confined by a symmetric harmonic potential, which becomes much larger that $k_B T$ only at large distances from the center of the funnel.
 
The specific functional form of the chosen potential energy function is 
\be
U({\bf x}) = U^F({\bf x}) + U^R({\bf x}) + U^H({\bf x}).
\ee 
$U^F({\bf x})$ is the term providing the funnel with an energetic barrier at its border and reads 
\be
U^F({\bf x})= a e^{-{\bf x}^2/(2 u^2)} - b e^{-{\bf x}^2/(2 v^2)},
\ee
where $a = 45\,kJ\,mol^{-1}$, $u = 25 nm$, $b = 45\,kJ\,mol^{-1}$ and $v = 10 nm$.
$U^R({\bf x})$ is the term providing the hump and reads 
\be
U_R({\bf x})= c\,e^{-({\bf x}-{\bf x}^0)^2/{2 w^2}}
\label{UR}
\ee
where ${\bf x}_0$ is the position of the corrugation with size  $w=3 nm$ and hight $c=12.5 kJ$. 

$U^H({\bf x})$ is a confining and symmetric harmonic trap,
\be
U^H({\bf x})=\frac{1}{2} \omega^2( {\bf x}^2),
\ee 
where  $\omega=0.11 kJ mol^{-1} nm^{-2}$. The resulting energy landscape is presented in Fig. \ref{landscape}.

With the protein folding problem in mind, we shall refer to the ensemble of configurations near the bottom of the funnel as to the native state. Conversely, the metastable flat region  between the harmonic trap and the energetic barrier near the entrance of the funnel  will be referred to as the denatured state. Note that the conformational entropy of the native state is much smaller than that of the denatured state, as it is the case for proteins. 

We are interested in studying the thermally activated transitions connecting an initial denatured configuration  $x_i$ to a native conformation $x_f$.
Due to presence of the hump, the landscape is not spherically symmetric and the structure of the relevant transition pathways depends on the temperature of the heat-bath. At asymptotically high temperatures, the diffusion becomes quasi-Brownian, as the reaction pathways are expected to be little sensitive to the local structure of the energy landscape. In such a temperature regime, the DRP approach is not expected to be reliable.
Conversely, in the low temperature regime, the diffusion will be driven by the structure of the force field: the most statistically relevant transitions will avoid visiting the positive corrugation and prefer to spend longer fractions of the transition time in the bottom of the funnel. 
In this regime the bundles of paths around different dominant reaction pathways will have negligible overlap and the DRP method will be reliable.
 \begin{figure}
\includegraphics[width=10 cm]{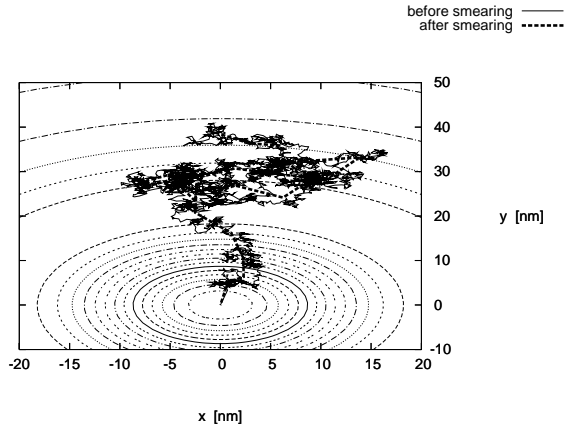}
	\caption{A typical starting trajectory before and after applying the smearing algorithm. }
\label{smearing}
\end{figure}

The main point of the DRP method is to first generate the set of {\it most probable} reaction trajectories  and then to use such information to efficiently sample the thermal fluctuations around each of such trajectories. This way one generates an ensemble of transition pathways, whose high statistical weight is guaranteed a priori, by construction. 

Any algorithm for the minimization of the HJ functional has to be provided with a starting path connecting the initial and final points, from which it begins the search for the most probable trajectories. 
In order to sample as efficiently as possible the space of transition pathways,  one needs to generate a full ensemble of {\it independent}  starting paths, each of which is defined in a relatively small number $N$ of discretization steps. 

In this simple example, we generate such an ensemble of initial paths as follows. We run a set of MD simulations starting from the same denatured conformation ${\bf x}_i$. The simulations  are stopped when the system reaches the native state, i.e. when the potential energy takes a value which is less than $2 k_B T$ larger than than the potential at the minimum of the funnel\footnote{Such a  prescription is suitable for the present  study, aiming to illustrate how the DRP works. However, implementing the same procedure becomes very computationally demanding, when applied to generate starting paths in complicated multi-dimensional systems. In the discussion of  the application of the DRP method to molecular systems, we shall adopt a different algorithm to generate the first path and choose a representative set of initial conformations \cite{Stoch2}.}.

The escape from the funnel represents thermally activated events, in which the microscopic time scale associated with the local motion in the native and denatured state is decoupled from the time scale associated to the inverse transition rate. Consequently, the initial trajectories obtained via  MD simulations contain a very large number of discretization steps $N_{MD}$. On the other hand, the main advantage of the DRP method is that it allows to keep only a small number of path steps $N_{DRP}$, typically of the order of $50-100$, for realistic molecular systems. Consequently, before beginning the minimization procedure, one needs to reduce the number of path steps in the initial MD trajectory from $N_{MD}$  down to $N_{DRP}\ll\,N_{MD}$. 
\begin{figure}[b]
\includegraphics[width=6 cm]{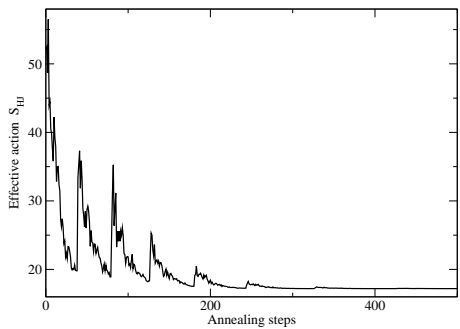}
	\caption{Typical evolution of the HJ action during the annealing algorithm. }
\label{anneal}
\end{figure}

In order to do so, we applied the following smearing algorithm to the initial path $\{{\bf x}(i)\}_{i=1...N_{MD}}$. At each iteration, we defined a new path $\{{\bf x}'(i)\}_{i=1...N_{MD}}$, consisting of $N_{MD}/2$ steps, in which each configuration was defined as average of two configurations of the input path at consecutive time steps, i.e. 
\be
{\bf x}'(i)=\frac{1}{2}\left({\bf x}(2 i +1)+{\bf x}(2 i) \right).
\ee
After $k$ of such decimation steps, one obtains a smeared trajectory consisting of $N_{DRP}=2^{-k} N_{MD}$ steps.
The shape of a typical initial trajectory  before and after smearing is reported in Fig. \ref{smearing}.
 The smearing algorithm  was used to reduce the number of steps in the trajectory from $\mathcal{O}(10^4)$  to $30-50$.

The minimization of the discretized HJ effective action was performed applying the following  simple  simulated annealing algorithm. After a preliminary thermalization phase, based on the Metropolis algorithm, we performed 10 cooling cycles, which were stopped when the acceptance rate dropped below $10\%$. After each cycle we performed a re-heating cycle consisting of few Metropolis steps. 
At each step we updated the path globally, i.e. moving the position of the particle in each configuration of the path. At the end of each cooling cycle, the boldness of the Monte Carlo moves was adjusted, in order to keep the acceptance rate between $10\%$ and $90\%$.  The evolution of the HJ action during the annealing of one of the initial trajectories is reported in Fig.  \ref{anneal}. For such a simple problem, convergence was typically achieved in few minutes on a single desktop.

Once a dominant trajectory $\ox(n)$, 
with $n=1,..,N_{DRP}$ has been determined by minimizing numerically the HJ action (\ref{DSHJ}) it is possible to sample the space of small thermal fluctuations around it by means of the Monte Carlo algorithm described in section \ref{SQ}, i.e. by sampling the
path integral (\ref{DPI}) with a simple Monte Carlo simulation, using the dominant reaction pathway determined before as the starting configuration of the Markov chain. Given the simplicity of the system under consideration, it was sufficient to adopt a Metropolis algorithm.

Let us now discuss the results concerning  the diffusion from the denatured configuration $ x_i=(0,50)[nm]$ to the native conformation $x_f=(0,0)[nm]$, in two different energy landscapes. In the first landscape, which we shall refer to as  $A$, the hump in the flat region was removed and the diffusion took place in a smooth symmetric funnel. In the energy landscape $B$, a hump of $ 5 k_B T$ at $T=300 K$ was placed between the native and denatured conformations. 
The dominant reaction pathway for such a reaction have been found at different temperatures, $T=50 K$ and $T=300 K$, by minimizing the HJ action using 10 independent initial trajectories.

\begin{figure}		
\includegraphics[width=8.5cm]{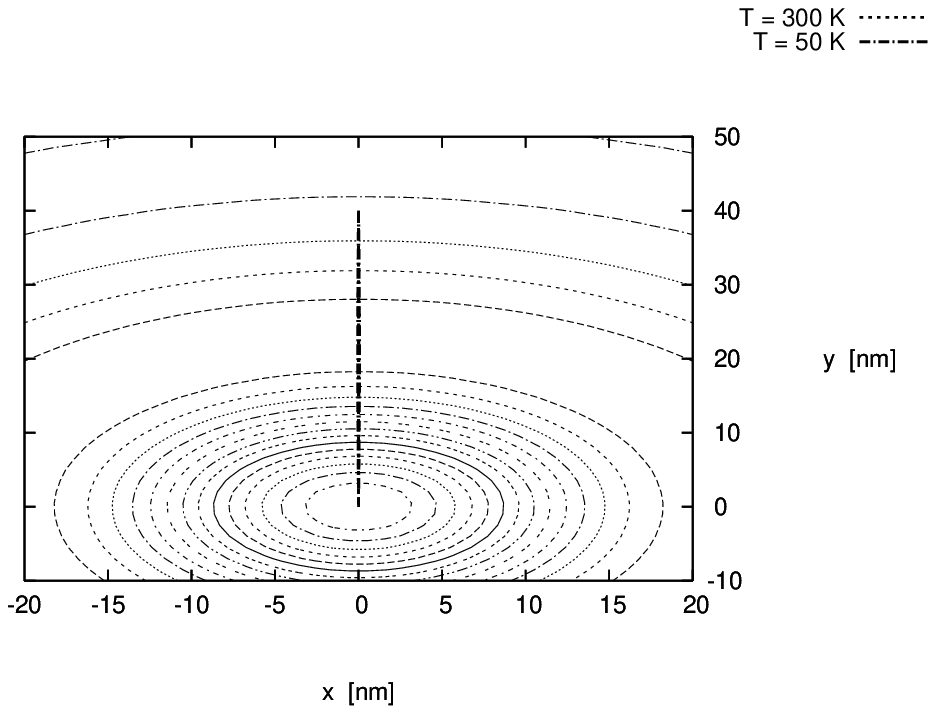}\hspace{0.5 cm}
\includegraphics[clip=,width=8.5cm] {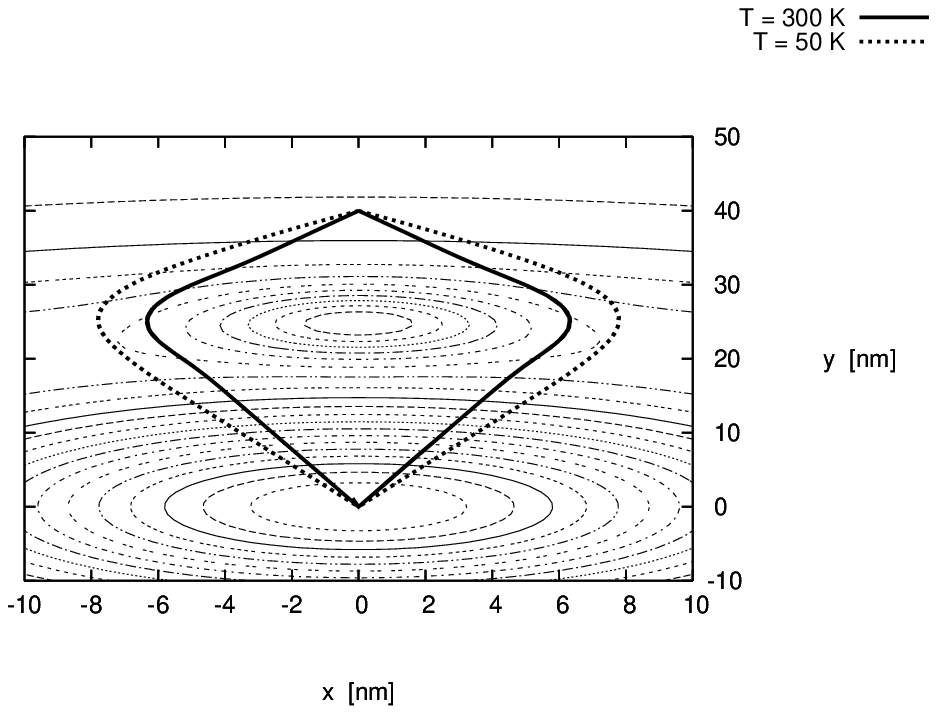}
	\caption{Left Panel: dominant reaction pathway for the diffusion in the landscape A, obtained from 10 different starting trajectories at $300 K$ (solid line) and $50 K$ (dashed lines). Right Panel: dominant reaction pathways for the diffusion in the landscape B obtained from 10 different trajectories at $50 K$ (solid line) and $50 K$ (dashed line).}
\label{scA}
	\end{figure}

In the  landscape $A$,  the minimization algorithm always converged to the same dominant path, regardless of the temperature of the heat-bath and on the trajectory used as  starting point of the annealing Markov chain  --- see the left panel of Fig. \ref{scA} ---.
Hence, for this transition, there is only one type of dominant path\footnote{In the present analysis, we are not concerned with multi-instanton paths which escape from and return to the bottom of the funnel, several times.}. 

In landscape $B$,  the minimizations always converged to two different dominant reaction pathways: one passing on the right and one passing  on the left of the barrier as shown in the right panel of Fig. \ref{scA}.
Notice that the two dominant paths tend to converge, as the temperature is increased.
This is well understood since, in the asymptotically  large temperature limit, the structure of the landscape becomes irrelevant and the diffusion becomes purely Brownian.
\begin{figure}[b]
\includegraphics[clip=,width=6.0 cm] {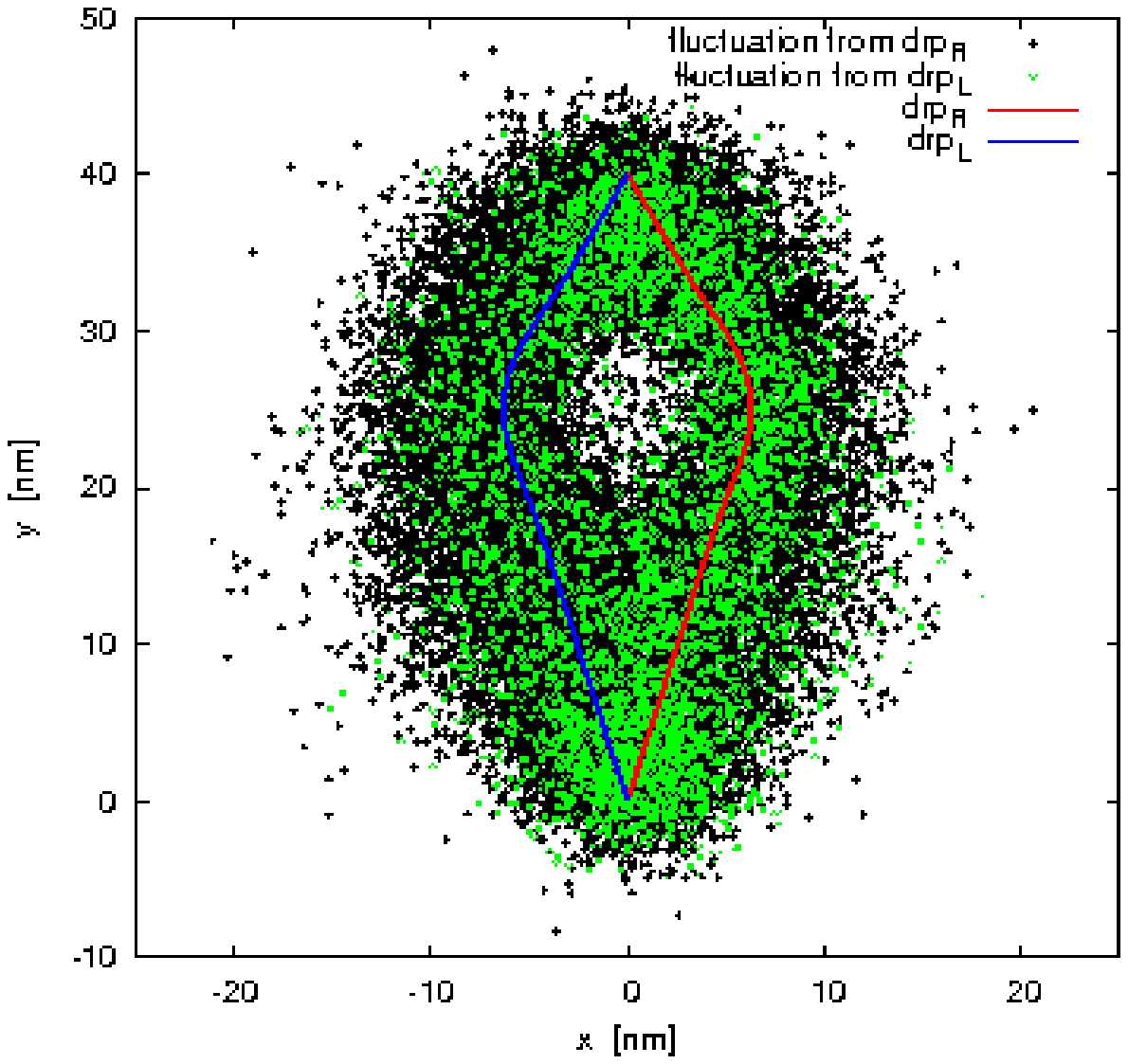}\hspace{2.5 cm}
\includegraphics[clip=,width=6.0 cm] {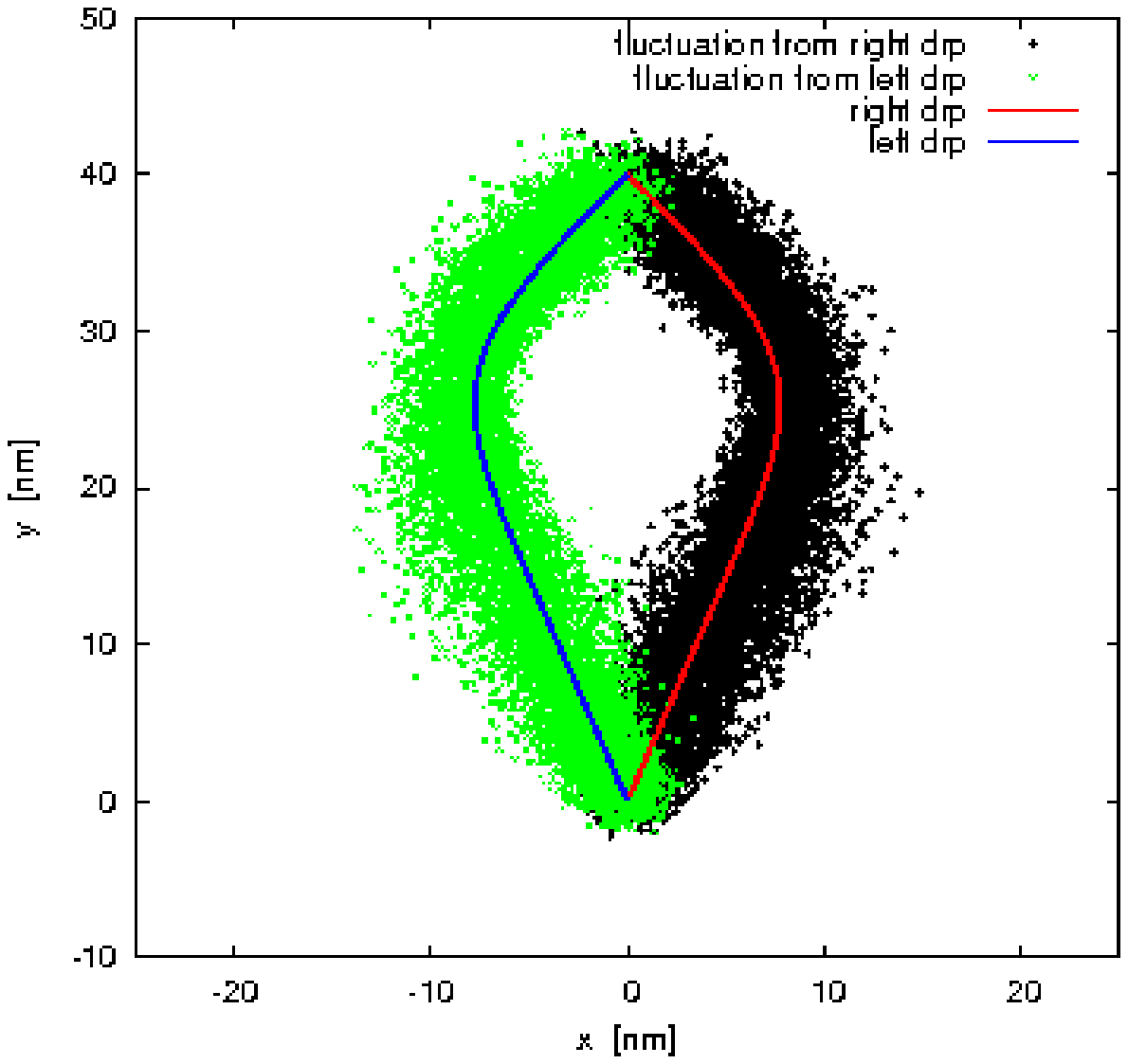}
	\caption{Left Panel: Ensemble of statistical fluctuations around the dominant reaction pathways at $T=300 K$. Right Panel: Ensemble of statistical fluctuations around the 
dominant reaction pathways at $T=50 K$.}
\label{scBT}
\end{figure}

The temperature of the heat bath does not only affect the structure of the dominant reaction pathways. Most importantly, it   governs the size of thermal fluctuations around the dominant pathways.
The panels of Fig. \ref{scBT} show the ensemble of statistically relevant fluctuations around the two dominant paths, obtained from the Monte Carlo sampling of the integral (\ref{DPI}),  at  temperature $T= 300 K$ and $T=50 K$, respectively. 
As expected, at low temperatures, the fluctuations are small and the bundles of paths associated with the two dominant paths do not overlap. 
Consistently, the Monte Carlo sampling performed starting from one of the two dominant paths takes an exponentially large time to generate paths which  cross the hump and explore the configurations in the other side of the valley. 
In this regime, it makes sense to distinguish between the two possible ways in which the transition can take place and the correct average trajectory is recovered only after one performs the average over the paths sampled around each dominant reaction pathways.
 
On the other hand, as the temperature is increased and becomes of the order of the height of the barrier, the two sets of fluctuation trajectories begin to overlap and eventually become indistinguishable. This is evident from the second panel of Fig. \ref{scBT}.
Notice that in all cases the DRP approach leads to the correct average trajectory which travels through the barrier in the middle. 

\begin{figure}
		\includegraphics[width= 8 cm] {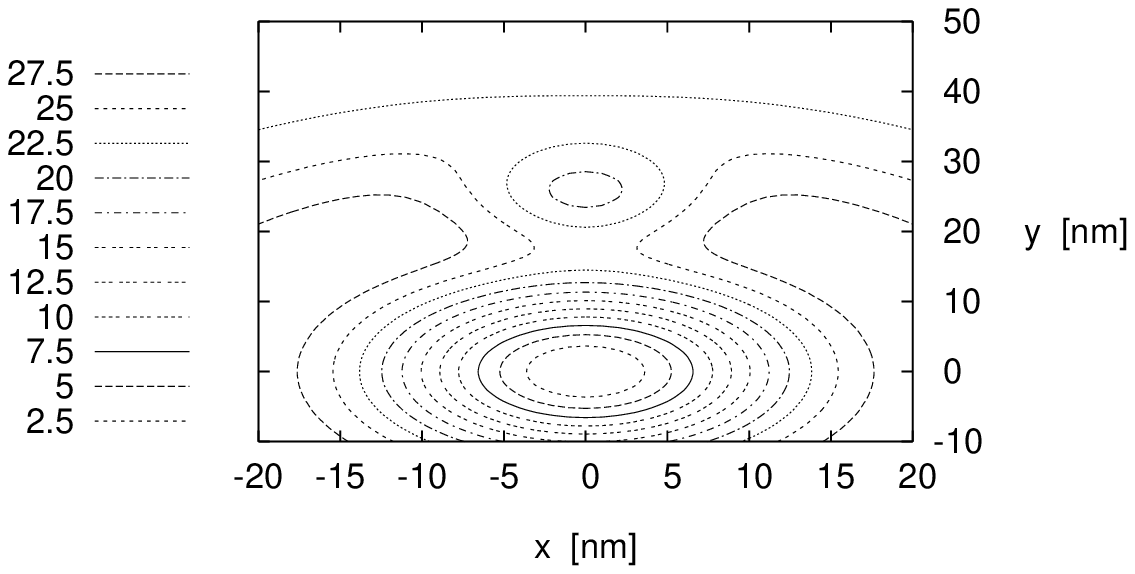}\hspace{1cm}
		\includegraphics[width=6 cm] {deltat.eps}
	\caption{Left panel: energy landscape C. Right panel:  residence time along the corresponding dominant reaction pathway. }
\label{restime}
\end{figure}

Let us now discuss the residence time in each of the configurations of a dominant reaction pathway.
To this end, we consider a third landscape $C$  which displays a negative depression, before the entrance of the funnel (left panel of Fig. \ref{restime}), corresponding to setting $c= 10 kJ$ and $w= 5 nm$ in Eq. (\ref{UR}). In the right panel of Fig. \ref{restime}, we plot the corresponding residence time along the dominant trajectory, as a function of the curvilinear abscissa $l$, which parametrizes the path. We clearly observe the presence of a local maxima of the residence time, when the particle travels near the depression. As expected, the residence time diverges  in the native state, because of our choice of effective energy $E_{eff}=-V_{eff}(x_f)$.

\section{Conclusions and Outlook}
\label{conclusions}

In this work, we have provided a  presentation of a theoretical framework to characterize the important thermally activated reaction pathways in multi-dimensional systems obeying Langevin dynamics. 
We have shown that, by applying the instanton  calculus to the path integral formulation of the stochastic  dynamics, it is possible to define, determine and characterize not only the most probable pathways, but also the  ensemble of most statistically significant thermal fluctuations around them. 

From the theoretical point of view, our method has the advantage that it does not require any {\it prior} choice of reaction coordinate. However, a natural reaction coordinate, $l$ emerges naturally from the HJ formulation of the dynamics of the dominant reaction pathways. In addition, this framework yields a simple procedure to generate  sets of conformation which are representative of the transition state.

From the computational point of view, the method has the advantage that it avoids investing CPU time in simulating the dynamics of the system in the metastable states, i.e. when it is not performing a transition to the final state.

As a result, rather than simulating long Molecular Dynamics (MD) trajectories until a significant number of transitions occur, we suggest to first determine the most probable transitions  and then to sample  the most significant fluctuations around them.  

\appendix
\section{Long time limit of the out-of-equilibrium average}
\label{av}

Here we show that, if  the average (\ref{Odef}) is evaluated in the  long time limit --- i.e. for  $t'\to (t/2)$ and $t\to \infty$---  it coincides with the equilibrium Boltzmann  average,
\be
\lim_{t'\to t/2, t \to \infty}   \langle O(t')\rangle =  \frac{\int dx e^{-\frac{U(x)}{k_B T}}\, O(x)}
{\int d x e^{-\frac{U(x)}{k_B T}}}.
\ee

We start from 
\be
\lim_{t'\to t/2} 
\langle O(t') \rangle =
\frac{\int dx e^{-\frac{U(x)-U(x_0)}{k_B T}} \langle x | \hat{O} \,e^{-\hat{H}_{eff} t} 
| x_0\rangle}{ 
\int dx e^{-\frac{U(x)-U(x_0)}{k_B T}} \langle x |\,e^{-\hat{H}_{eff} t} 
| x_0\rangle }.
\label{Odef2}
\ee
We now insert a resolution of the identity in terms of a complete set of eigenstates of the effective quantum Hamiltonian $|n\rangle$:
\be
\langle O(t) \rangle &=& 
\frac{ \int  dx  \sum_{n}  e^{-\frac{U(x)-U(x_0)}{2 k_B T}} \langle x|  \hat{O} \,e^{-\hat{H}_{eff} t} 
| n\rangle \langle n | x_0\rangle}{ 
 \int dx ~ \sum_n e^{-\frac{U(x)-U(x_0)}{2 k_B T}} \langle x |\,e^{-\hat{H}_{eff} t} 
|n\rangle \langle n  | x_0\rangle }.\nonumber\\
&=&
\frac{  \int dx~ \sum_{n} \,e^{-\frac{U(x)-U(x_0)}{2 k_B T}} e^{-E_{n} t} O(x)  
\psi^\dagger_n(x) \psi_n(x_0)}{ 
\int dx~ \sum_n e^{-E_n t }\ e^{-\frac{U(x)-U(x_0)}{2 k_B T}} \psi^\dagger_n(x) \psi_n(x_0)},
\ee
where we have introduced the real "wave functions" $\psi_n(x)\equiv \langle n | x \rangle = \psi^\dagger_n (x)\equiv \langle  x | n\rangle$.

In the long time limit, the contributions of all the eigenstates $| n\rangle $, except the ground-state are exponentially suppressed and one gets
\be
\langle O(t) \rangle =  \frac{\int dx e^{-\frac{U(x)-U(x_0)}{ 2 k_B T}}\, O (x) e^{-E_0 t} \psi_0(x_0) \psi_0^\dagger(x)}{\int dx e^{-\frac{U(x)-U(x_0)}{ 2 k_B T}}\,  
e^{-E_0 t} \psi^\dagger_0(x)  \psi_{0}(x_0)} .
\ee
We now recall that for any systems obeying the  Fokker-Planck Eq.(\ref{FPE}), the ground-state of the effective Hamiltonian is  $\psi_0(x)=e^{-\frac{U(x_0)}{2 k_B T}}$ and  the corresponding energy is null. 
This fact immediately yields Eq.  (\ref{Obol}). 

\section{Integration in the zero-mode space}
\label{dc0}

In the presence of barrier crossing pathways, the WKB expression (\ref{KtWKB}) breaks down and the integration over the zero mode coefficient $c_0$ cannot be performed in gaussian approximation, as
\be
\int_{-\infty}^{\infty} \frac{d c_0}{\sqrt{(2\pi)^d}} e^{-\frac{1}{2} \lambda_0 c_0^2} = \int_{-\infty}^{\infty} \frac{d c_0}{\sqrt{(2\pi)^d}}
\ee
Let us compare the integration the deviation $d x$ from a given path which are caused by
a small time translation of the time at which the transition takes place $t_0\to t_0 + d t_0$ with the deviation caused by a small coefficient shift, $c_0\to c_0 + d c_0$. The shift in $t_0$ leads to 
\be
d x = \frac{d\ox_i}{d t_0} d t_0 = - \dot{\ox_i} d t_0,
\ee
while the shift $c_0$ leads to
\be
d x = \frac{d\ox_i}{d c_0} d c_0 = \tilde{x}_0 d c_0 = \sqrt{\frac{1}{2D S_{eff}}} \dot{\ox_i} d c_0
\ee
Equating the two variations and redefining the sign of $c_0$ so that the two integrations have the same boundaries, we find
\be
d c_0 = \sqrt{2 D S_{eff}} d t_0.
\ee

\section{A simple analytic example}
\label{free}
Here we  illustrate the construction of the leading-order estimate of  the Fokker-Planck conditional probability $P(x_f t/2| x_i,-t/2)$ using the HJ formalism, in the simplest possible case, i.e. for  pure Brownian motion.

By rotational invariance,  the Fokker-Planck probability is  a function only of the distance $L=|x_f-x_i|$  and of the time interval $t$. To lowest-order in the saddle-point approximation we have
\be
P(L, t)\simeq e^{E_{eff}(t) t - S_{HJ}(E_{eff}(t),L)}.
\ee
The dominant reaction pathway for this transition is obviously the straight line path, since the  HJ action is proportional to the distance covered. Let us first evaluate the effective energy parameter  as a function of the time interval , $E_{eff}(t)$,
\be
t= \int_i^f dl \frac{1}{\sqrt{4 D E_{eff}(t)}}=  \frac{L}{\sqrt{4 D E_{eff}(t)}},
\ee
which gives
\be
E_{eff}(t)= \frac{L^2}{4 D t^2}. 
\ee

The HJ action is 
\be
S_{HJ}(E_{eff}(t), L)&=& \int_i^f dl \sqrt{\frac 1 D (E_{eff}(t))} =  L \sqrt{\frac{E_{eff}(t)}{D}}\nonumber\\
&=& \frac{L^2}{2 D t}
\ee

We are therefore able to reconstruct the Fokker-Planck probability 
\be
P(L,t)\simeq   {\textrm const.} \cdot e^{\frac{L^2}{4 D t}-\frac{L^2}{2 D t}} = {\textrm const.} \cdot e^{-\frac{L^2}{4 D t}}
\ee
For such a simple problem the leading-order approximation yields the exact result. The normalization of the probability is of course given by the quadratic fluctuations.

\end{document}